\input{epsf}

\documentclass[preprint,showpacs,preprintnumbers,amsmath,amssymb]{revtex4}

\usepackage{graphicx}

\begin{document}
\draft
Published as: Astropart. Phys., 41 (2013) 1–6
\title{Tachyonic neutrinos and the neutrino masses}

\author{Robert Ehrlich}
\address{George Mason University}

\begin{abstract}
With a recent claim of superluminal neutrinos shown to be in error, 2012 may not be a propitious time to consider the evidence that one or more neutrinos may indeed be tachyons.   Nevertheless, there are a growing number of observations that continue to suggest this possibility -- albeit with an $m_{\nu}^2<0$ having a much smaller magnitude than was implied by the original OPERA claim.  One recently published non-standard analysis of SN 1987A neutrinos supports a tachyonic mass eigenstate, and here we show how it leads to 3 + 3 mirror neutrino model having an unconventional mass hierarchy.  The model incorporates one superluminal active-sterile neutrino pair, and it is testable in numerous ways, including making a surprising prediction about an unpublished aspect of the SN 1987 A neutrinos.   Additional supporting evidence involving earlier analyses of cosmic rays is summarized to add credence to the tachyonic neutrino hypothesis.
\end{abstract}

\pacs{13.15+g, 14.60Pq, 14.60St, 14.60Lm}

\maketitle

\section{Introduction}

In this paper we summarize various observations suggesting that one of the neutrinos is consistent with being a tachyon as originally defined, i.e., a particle with $m^2<0$ and $v > c$ that obeys relativistic kinematics,\cite{Bilaniuk} a possibility first raised by Chodos, Hauser and Kostelecky.\cite{Chodos}   As is well known, time-of-flight measurements of neutrinos no longer show any indication of superluminality, but they do set useful upper limits at the GeV energy scale.\cite{Bertolucci, Adamson}   There is also the upper limit on $\delta=(v-c)/c$ at low energies (around 20 MeV) set by SN 1987A, i.e., $\delta < 2 \times 10^{-9}$.\cite{Kamioka}  As shown in the next section, however, there are reasons to disbelieve this much more stringent upper limit.   

\section{SN 1987A neutrino data}

\subsection{Questioning the upper limit on $\delta$}

The burst of 24 neutrinos seen in the Kamioka,\cite{Kamioka} IMB\cite{IMB} and Baksan\cite{Baksan} detectors, arrived about 3 hours before the light was recorded from SN 1987A.  This early arrival was presumably due to the delay experienced by photons emitted from the collapsing SN core, which was not the case for the emitted neutrinos.  However, the value of the photon delay need not have been the entire 3 h, therefore the early neutrino arrival is normally assumed to set only an upper limit on any excess above c for their speed, $\delta < 2 \times 10^{-9}$.  Here we show that one cannot rule out a third superluminal mass eigenstate that arrived long before the other 24 neutrinos.  This assertion, however, does not refer to the burst of 5 events observed in the LSD detector underneath Mt. Blanc,\cite{Mt Blanc}  which occurred during a 7 s interval nearly 4 hours before the 24 event burst, as we can easily show.  Consider hypothetical superluminal neutrinos of some fixed $m^2$ and varying values for their energy $E$ that are assumed to have originated in a very brief burst.  Relativistic kinematics under the approximation that $\sqrt{1-m^2/E^2} \approx 1-\frac{m^2}{2E^2}$ requires that the neutrino arrival time t can be expressed as 

\begin{equation}
t = \frac{m^2t_0}{2E^2}
\end{equation}

where $t_0$ denotes the light travel time from the supernova, and $t=0$ would be the arrival time of $m^2 = 0$ neutrinos -- which as shown in reference 12 is probably equivalent (within $\pm 0.5 s$) to setting $t = 0$ for the earliest arriving neutrino in each detector for the 24 event burst.  From Eq. 1 we therefore find that if $m^2$ is fixed that the spread in the neutrino arrival times will be related to the spread in their energies according to:

\begin{equation}
\frac{\Delta E}{E}=\frac{\Delta t}{2t}=\frac{7 s}{2\times 4 h}=\frac{1}{5140}
\end{equation}
  
Eq. 2 implies that in order to be observed $t=4h$ early within a burst as short as $\Delta t =7 s$ the superluminal neutrinos would have had to be monochromatic to one part in 5140 -- which is virtually inconceivable for neutrinos from an exploding supernova.  Turning the argument around, we can say that superluminal neutrinos with the energy spread seen for events in the three detectors, i.e., $\frac{\Delta E}{E} \approx 1$ would have arrival times spread over many hours and would certainly not be recognized as a pulse above background (around 1 event in 8 seconds).   Whatever the source of the Mt. Blanc neutrinos, they could not have been due to brief superluminal burst emitted from SN 1987A.  The inability to recognize a superluminal signal as a short pulse above background would be even less possible for larger excesses above light speed, where the spread in arrival times would be even larger.  Thus,  the normally assumed upper limit $\delta < 2 \times 10^{-9}$ from SN 1987A data is not correct, because any real superluminal signal would have gone unnoticed for large $\delta$ if one is expecting to see a pulse above background.

\subsection{Two claimed mass eigenstates}

The neutrinos from SN 1987A have been the subject of hundreds of papers, both theoretical and phenomenological.\cite{Pagliaroli}  Some of these papers analyze the data to infer an upper limit on the electron neutrino mass, which ranges typically from 12 to 16 $eV$,\cite{Bahcall, Arnett} although one 2010 analysis has claimed a 5.8 eV upper limit,\cite{Pagliaroli1} and still more refined methods may allow future galactic supernova to achieve mass limits as low as 0.14 eV.\cite{Ellis}  In marked contrast to finding upper limits, a 2012 paper has claimed evidence for the presence of two (non-superluminal) mass eigenstates for the SN 1987 A neutrinos,\cite{Ehrlich0} following the method of earlier similar analyses by Cowsik\cite{Cowsik} and Huzita.\cite{Huzita} The heavier mass eigenstate has  $m_2=21.4 \pm 1.2 eV$, while the lighter one has  $m_1=4.0 \pm0.5 eV$ -- similar to the values cited by Cowsik in 1988,\cite{Cowsik} but with considerably smaller uncertainties.    

Before considering the implications of this result for a third superluminal state having $m_3^2<0,$ and why such outlandishly heavy mass eigenstates need not conflict with well-established upper limits on neutrino masses, e.g., from cosmology, a brief summary of the basis of the claim of 4.0 and 21.4 eV mass eigenstates is in order.  The analysis is based on an observed correlation between recorded  neutrino energies $E_k$ and arrival times $t_k$ for the $k = 1, 2, \cdots 24$ events in the three detectors (excluding the 5 events from the Mt. Blanc detector).  If the neutrinos reaching Earth were all emitted nearly simultaneously then based on Eq. 1 on a plot of $1/E^2$ versus $t$ all those neutrinos having a mass $m_1$ will lie on a line of slope $2/(t_0m_1^2)$ while those having a mass $m_2$ will lie on a line of slope $2/(t_0m_2^2).$  Fig. 1 of reference 12 clearly shows that every one of the 24 neutrinos do lie on or near one of two straight lines, and the two fits have acceptable chi square.  The fact that those two straight lines also nearly pass through the origin implies that the choice of $t=0$ for the first arriving neutrino in each detector made by each experiment nearly conforms to the definition of $t=0$ used in Eq. 1.  It should also be noted that the claim of two mass eigenstates is not contradicted by the fact that all arriving neutrinos are detected only as (anti)electron neutrinos, because it is only for mass not flavor eigenstates that all neutrinos having some specific energy E travel at some fixed mass-dependent speed.   Moreover, supernova neutrino data is unique in that no other time of flight measurement could possibly have the time resolution to observe separate mass eigenstates, since the distance to SN 1987A is approximately $3 \times 10^{14}$ Earth diameters.

The main weakness of the claim of two mass eigenstates is that it rests on there being near-simultaneous supernova neutrino emissions (within $\pm 0.5 s$) of most of the detected SN 1987A neutrinos.  Supernova core collapse models in fact do show that the burst of electron neutrinos and antineutrinos rises and falls by an order of magnitude in the first second,\cite{Totani, Bruenn} while some models show it lasting only about 0.02 seconds.\cite{Myra} Alternatively, it is possible some of the neutrinos detected from SN 1987A were emitted over an extended period of time, but they had a strange correlation between their energy and emission time that mimicked two mass eigenstates on an plot of $1/E^2$ versus $t$.  One could conceivably accommodate this correlation within the framework of a composite model consisting of the sum of two thermal spectra.\cite{Lamb}  Ultimately, however, there is no way to know precisely what fraction of the neutrinos emitted during a supernova core collapse are emitted in the first second.  While supernova modeler Thomas Janka has suggested the number is likely to be no more than half,\cite{Janka} the fraction of the 24 $\emph{observed}$ neutrinos emitted during the first second could be considerably greater than half, given the softer spectra of later-emitted neutrinos.\cite{Raffelt}

\subsection{SN 1987A and superluminal neutrinos?}
In the remainder of this section we show that even though the two mass eigenstates claimed for SN 1987A are not superluminal their existence (if confirmed) would imply that there must be a third unobserved eigenstate that is superluminal in order to be compatible with cosmological upper limits on the sum of the masses of the three flavor states, i.e., $\sum^3_{j=1} |m_j| < 0.28 eV,$\cite{Thomas} and that of the electron neutrino mass, ${m_\nu}_e < 2 eV$ from tritium beta decay.\cite{PDG}  

We can express the effective mass of the fth flavor state in terms of a sum over all the mass eigenstates $m_i$ as:

\begin{equation}
m^2_f=\Sigma_i|U_{f,i}|^2 m^2_i
\end{equation}

Thus, in light of the large values of $m_1$ and $m_2$, only if the 3rd mass eigenstate has ${m_3}^2 <0$ could the three flavor eigenstates all be quite close to zero, having either $m_f^2>0$ or $m_f^2<0,$ depending on the values of the $U_{f,i}$.  Equivalently, the known upper limits on $\sum^3_{j=1} |m_j|$ and ${m_\nu}_e$ together with the confirmed existence of $4.0 eV$ and $21.4 eV$ neutrino mass eigenstates would \emph{require} that the third mass eigenstate be superluminal with $m_3^2<0$.

\section{A 3+3 neutrino model and the neutrino mass hierarchy}

\subsection{How many sterile neutrinos?}

One might also object that masses as large as 4.0 and 21.4 eV are incompatible with the two well-measured neutrino oscillation results for the $\Delta m^2$: $7.6 \times 10^{-5} eV^2$ and $2.4 \times 10^{-3} eV^2$ that suggest very small values for the masses themselves.  However, one could accommodate those two measured values if there were three sterile neutrinos two of which were nearly degenerate with the 4.0 eV and 21.4 eV active neutrinos, and also if the measured $\Delta m^2$ values to date are between one active and one sterile neutrino.   The empirical basis for sterile neutrinos has been gaining strength, since the entire collection of neutrino oscillation experiments can be fit with one or more sterile neutrinos, and they fit best with three of them.\cite{Kopp, Conrad}   For example, a collective fit with one sterile neutrino has a good probability of 55\% but a compatibility between data sets of only 0.043\%, while the three sterile neutrino fit has a $90\%$ probability and a compatibility of $53\%$ between data sets.\cite{Conrad}   Although one might not be especially impressed with a good fit having as many as free parameters as occurs with 6 oscillating neutrinos, as we shall see later, the number of $\emph{independent}$ free parameters is far less than what one might think.

\subsection{A 3 + 3 neutrino model}

Here we discuss a 3 + 3 neutrino model assuming three active/sterile pairs, which differs significantly from earlier models,\cite{Berezhiani} because (a) one pair is superluminal, (b) the mass splitting of each active-sterile pair is very small, and (c) the $\Delta m^2$ seen in oscillation experiments to date are all between a sterile and an active neutrino.   The 3 + 3 model mass hierarchy is depicted in Fig. 1, and it consists of three right/left handed neutrino doublets, whose mass splittings for the two $m^2>0$ (tardyonic) doublets are taken to be the values found from neutrino oscillation experiments.   It is of course the right handed states that are the sterile ones -- at least in the case of the two tardyonic doublets.  It should be noted that if the $4.0 eV$ and $21.4 eV$ mass eigenstate result is genuine, the mass hierarchy suggested in Fig. 1 is the only one compatible with neutrino oscillation experiments, with the exception of switching the two labeled $\Delta m^2$ values and reversing the order of each R and L state.

It is extremely interesting that the two mass splittings labeled in Fig 1. when expressed as a fraction of each doublet's $m^2$, are identical within experimental uncertainties, i.e., $\frac{\Delta {m_1}^2}{{m_1}^2} = 4.8 \times 10^{-6}$ and $\frac{\Delta {m_2}^2}{{m_2}^2} = 5.2 \times10^{-6}$.   If the fractional mass splitting for the third (superluminal) mass doublet has the same value, and $\Delta {m_3}^2\approx 1 eV^2$ as suggested by short-distance accelerator and reactor neutrino experiments,\cite{Louis} we then find an approximate mass of the superluminal mass doublet: ${m_3}^2 =-\frac{\Delta {m_3}^2}{5 \times 10^{-6}} \approx -200,000 eV^2=-0.2 keV^2$.  

\subsection{Global fits to 3 + 3 models}

In a 3 + 3 global fit to all experiments the main interest is in seeing whether any large $\Delta m^2$ values are required to get a good fit (beyond the well-measured 3 small $\Delta m^2$), and what constraints can be placed on their values.  As is well known, in the standard mass hierarchy, one assumes that the three active neutrino masses have $m << 1 eV,$ and the one or more sterile neutrinos are considerably heavier.  The basis of this hierarchy rests on the assumption that the two very small well-measured $\Delta m^2$ values from oscillation experiments are both between the three active neutrinos (1,2,3), while in our model they are each between the two members of an active-sterile doublet.  In other words, the well-measured value of $\Delta m_{1,2}^2$ is assigned to our $\Delta m_{1,4}^2$ and that of $\Delta m_{2,3}^2$  to our $\Delta m_{2,5}^2$ -- a difference that is irresolvable in a given experiment, whether one is doing a search for actual oscillations or the appearance/disappearance of a flavor state.   The distinction between the two model hierarchies is of vital importance, however, when doing a 3 + 3 global fit to all experiments, since it will affect the best fit values of the three large $\Delta m^2.$  Thus, the Conrad et. al. 3 + 3 fit was based on the standard mass hierarchy in which the three active neutrinos are much lighter than the three sterile ones.\cite{Conrad}  In this fit there are only three independent large $\Delta m^2$ values which they take to be $\Delta m_{14}^2, \Delta m_{15}^2$ and $\Delta m_{16}^2.$   Only the first of Conrad's three fitted large $\Delta m^2$ agree with those predicted from our 3 + 3 model, which are $1.0eV^2,$ $200,000eV^2,$ and $21.4^2-4.0^2=442 eV^2$ -- see Fig. 1. This disagreement is to be expected, of course, given the differences between our mass hierarchy and the conventional one Conrad et. al. used in doing their fit.  Thus, in our 3 + 3 model there is only one $\Delta m^2=2.4\times 10^{-3} eV^2$ not two.  In addition the three independent large $\Delta m^2$ values in our model are $\Delta m_{12}^2, \Delta m_{13}^2$ and $\Delta m_{36}^2$ (3 double arrows in Fig. 1), and the other large $\Delta m^2$ are given in terms of them.  For example, these eight large $\Delta m^2$ are all approximately equal to three significant figures: $\Delta m_{13}^2, \Delta m_{34}^2,  \Delta m_{46}^2, \Delta m_{16}^2, \Delta m_{23}^2, \Delta m_{26}^2, \Delta m_{35}^2, \Delta m_{56}^2,$ as are these four: $\Delta m_{12}^2, \Delta m_{15}^2, \Delta m_{24}^2, \Delta m_{45}^2.$  These equivalencies would be more obvious had all the $\Delta m^2$ spacings been drawn to scale.

\section{A digression: Earlier work suggesting ${m^2_\nu}_e \approx - 0.16 eV^2$}

Given the highly controversial nature of the claims made in the previous section it is useful to summarize earlier evidence for tachyonic neutrinos before considering theoretical support for them, and how those claims can be tested -- in some cases using existing data.

\subsection{The shape of the high energy cosmic ray spectrum}

Chodos, Hauser and Kostelecky in 1985\cite{Chodos} suggested that one could test whether electron neutrinos are tachyons based on the beta decay of stable particles whose energy exceeds some threshold.  In 1999, following a suggestion by Kostelecky,\cite{Kostelecky} Ehrlich adopted this idea to modeling the cosmic ray spectrum.\cite{Ehrlich1, Ehrlich2}  It is well known that the observed spectrum satisfies a power law $\frac{dN}{dE}\approx E^{-\gamma}$ where $\gamma$ changes value relatively abruptly at an energy in the vicinity of 4 PeV, which is known as the knee of the spectrum.  One can interpret the presence of the knee using the Chodos et. al. idea that protons are decaying with this energy as their threshold, and they are increasingly depleted from the spectrum above this energy.  For protons, the threshold energy is inversely related to the absolute value of the tachyon mass (in $eV$) through\cite{Ehrlich1}

\begin{equation}
E_{th}=\frac{1.7 PeV}{\sqrt{-m_{\nu_e}^2}}
\end{equation}

A second change in the spectrum power law known as the ankle occurs around $10^4 PeV.$   In Ehrlich's model a good fit was obtained to the high energy spectrum (including both the knee and the ankle), by assuming ${m_{\nu _e}}^2= -0.16\pm .09 eV^2.$ 

\subsection{Neutral hadrons in the cosmic rays from Cygnus X-3}

One important prediction of the fit to the cosmic ray spectrum was the existence of a neutron line (from proton decay) that occurred right at the knee.\cite{Ehrlich1}  Evidence for a neutron line at the knee was subsequently reported based on cosmic rays pointing back to Cygnus X-3, an X-ray binary having a 4.79 h period.\cite{Ehrlich3}   At PeV-scale energies cosmic rays pointing back to a particular distant source constitutes evidence that those primary cosmic rays are neutral particles, given the strength of the galactic magnetic field.  Of the four groups that had reported high energy cosmic rays from Cygnus X-3 during the 1970$'$s and 1980$'$s with signal strengths at the 4-5 $\sigma$ level, only the Lloyd-Evans group had events above 1 PeV.\cite{r14}    That data showed an excess of 28.4 events in two adjacent energy bins straddling 5 PeV, with an uncertainty in the two bin total of 5.0 events, i.e., $N=28.4\pm5.0$ ($5.7\sigma$).  Thus, the energy at which the peak occurs came very near the knee of the spectrum, which was previously interpreted as the threshold for protons to beta decay into neutrons.  Given that evidence also existed for the Cygnus X-3 cosmic rays being neutral hadrons, Ehrlich interpreted the 4.5 PeV peak associated with Cygnus X-3 as a confirmation of the earlier prediction of a neutron line at the knee of the cosmic ray spectrum.\cite{Ehrlich3}   Today many cosmic ray researchers express skepticism about the reality of those early reports of cosmic rays from Cygnus X-3.  The conventional wisdom is that the only primary cosmic rays pointing back to sources at PeV energies are photons or neutrinos.   In fact, a more recent high statistics cosmic ray study failed to observe non-episodic cosmic rays from Cygnus X-3 at PeV energies.\cite{r13}   However, it should be noted that this negative result need not disprove the validity of the earlier observations since Cygnus X-3 is known to be an episodic source that is especially intense at times of strong radio flares when the RF luminosity increases a thousand fold.  

\section{Theoretical support and challenges to $m^2<0$ neutrinos}

While there is yet no commonly accepted field theory of tachyons, a number of researchers including Ciborowski, Rembielinski, and Radzikowski have made important steps towards such a theory.\cite{Ciborowski, Radzikowski}   There are, however, many theoretical reasons that have been cited for being skeptical of superluminal neutrinos, including the magnitude of the violation of Lorentz Invariance (VLI) that they might entail.\cite{Cowsik1}  However, while these VLI constraints and the Cohen-Glashow (C-G) effect\cite{Cohen} conflict with a $\delta$ as large as the initial spurious OPERA value, they do not rule out much smaller values below the upper limits set by various experiments.\cite{Miniboone, Adamson1, Icecube}  In particular, note that for the C-G effect the threshold energy for tachyonic neutrino bremsstrahlung varies as $\delta^{-15},$  so the effect is of no consequence for very small $\delta.$  Thus, neither VLI nor the C-G effect is an argument against superluminal neutrinos in general.  Moreover, not only has VLI been shown to be compatible with extensions of the standard model,\cite{Kostelecky2} but Chodos recently has provided a theoretical rationale for tachyonic neutrinos.\cite{Chodos2}  He shows that one can construct a Lagrangian that satifies Light Cone Reflection, a new spacetime symmetry that arises in the context of a limited form of Lorentz Invariance, in which $\pm m^2$ neutrino pairs arise naturally.   Unfortunately, since Chodos' model explicitly requires that  $m^2_{tachyon} = -m^2_{tardyon},$ it is not consistent with our 3 + 3 model.  One that is consistent is a theory by Jentschura and Wundt who generalize the Dirac equation, based on a pseudo-Hermitian Hamiltonian. Their theory allows for a left-handed tachyonic neutrino mass eigenstate with a free mass parameter that is compatible with our value of $m_3^2 = -200,000eV^2,$\cite{Jentschura} and it also leaves room for the addition of sterile neutrinos.\cite{Jentschura2}  Moreover, the same authors note that these extensions to the Dirac equation allow tachyonic neutrinos to be a candidate for the acceleration of the universe or dark energy.\cite{Jentschura1} 

\section{Tests of tachyonic neutrinos and the 3 + 3 model}

The first proposed test is the least intereting one, and is likely only to result in setting upper limits, while all the rest involve testing far more specific predictions discussed earlier.

\subsection{Time of flight experimental searches for no specific ${m_\nu}^2<0$}

Time of flight experiments involving Earthly distances should be feasible so long as a tachyon had an $-m^2$ on the order of many $keV^2.$   The lowest mass one might be able to detect using Earthly distances might be around $m^2= - 0.0019 MeV^2= - 1,900 keV^2$ based on Eq. 1 using $E=20 MeV$, $t = 50 ns$ and $t_0 = 6400 km/(3\times 10^5 km/s)$.  Such a result would be many orders of magnitude smaller than the original OPERA claim.

\subsection{Confirming that ${m^2_\nu}_e  \approx -0.16 eV^2$}

If the forthcoming Katrin experiment designed to have a sensitivity of ${m_\nu}_e  = 0.2 eV$ or $|{m^2_\nu}_e|  = 0.04 eV^2$\cite{Katrin} fails to see a tachyonic electron neutrino, a PeV-scale cosmic ray experiment might well do so.  In particular, finding confirming evidence for cosmic ray protons decaying at or above the knee of the spectrum and giving rise to a neutron line at that energy would be very convincing evidence.  Of course, PeV-scale cosmic rays from the binary star Cygnus X-3  have been widely dismissed as an aberration by most cosmic ray researchers, in light of the very high statistics later negative search which found no indication of a steady state signal from that source.\cite{r13}  A new experiment that looked at Cygnus X-3 and other possible sources, but $\emph{only}$ at times of large flares, and in a narrow phase window based on the binary's period might well show a signal.

\subsection{Confirming the existence of mass eigenstates $m_{\nu_1}=4.0eV$ and $m_{\nu_2}=21.4eV$} 

An observation that confirmed the existence of these two mass eigenstates claimed based on the SN 1987A data, would by implication show the need for a third superluminal mass eigenstate in order to conform to existing mass limits on the sum of the flavor state masses.   One possibile test might come from a global fit to all oscillation experiments to see if one finds three large $\Delta m^2$ values, consistent with those predicted by the 3 + 3 model masses, as discussed in an earlier section.  Another test would be provided by the fortunate occurence of a supernova in a our galaxy or another one nearby, but these occur only about twice in a century.  There is, however, no need to wait for another galactic supernova, since the existence of mass eigenstates $m_{\nu_1}=4.0eV$ and $m_{\nu_2}=21.4eV$ are quite within the realm of a short baseline neutrino oscillation experiment.  For example, given $\Delta m^2= 21.4^2-4.0^2=442 eV^2,$ we find $\lambda=\frac{2\pi E}{1.267\Delta m^2}=11.2m$ (for E = 1 GeV).   A particularly suitable neutrino source might be the Spallation Neutron Source (SNS) at the Oak Ridge National Laboratory, given its high intensity, short pulse width, and the large percentage (about 30 \%) of neutrino flux that is monochromatic ($E=30 MeV$).  At this neutrino energy we find that $\Delta m^2= 442 eV^2$ would yield an oscillation wavelength of about 34 cm, which could be readily observed.

\subsection{Confirming that ${m_{\nu_3}}^2 \approx -0.2 keV^2$}

Searching for a ${m_{\nu_3}}^2 \approx -0.2 keV^2$ neutrino in an oscillation experiment should be doable at very high neutrino energies.  For example, the predicted oscillation wavelength between ${m_{\nu_3}}^2 \approx -0.2 keV^2$ and some much smaller mass is $\lambda=24.7 m$  (for E = 1 TeV) or $\lambda=2.47 m$  (for E = 100 GeV).   However, such a test could not distinguish between ${m_{\nu_3}}^2 \approx -0.2 keV^2$ and ${m_{\nu_3}}^2 \approx +0.2 keV^2.$  One measurement that could make this sign distinction would be an unreported and possibly unexamined aspect of the existing data on neutrinos from SN 1987A.  For a typical neutrino energy of 20 MeV,  $-0.2 keV^2$ neutrinos would have arrived around 25 min earlier than the main neutrino pulse.  Of course, based on Eq. 2 any such superluminal neutrinos would be spread over many minutes and would not be recognized as a pulse above background because of their spread in energy.  Nevertheless, given the energy dependence of the background events, there is a surprisingly simple way to discern a superluminal signal -- at least for the Kamioka data for which nearly all the background events in the detector have energies below 12 MeV -- the height of the dashed line in Fig. 2 (lower graph).  

The 17 minute time interval depicted in Fig. 2 (lower) includes the 12 event neutrino burst reported by Kamioka seen just after 7:35 UT.   In order to investigate background more thoroughly Kamioka has provided similar plots for 7 other time intervals selected at random, some before the 12 event pulse and some after.\cite{Kamioka}  If we exclude the one 17 minute interval that happens to fall in the one hour before the 12 event burst, Kamioka shows only one background event out of about 1000 that has an energy above 12 MeV in the entire 7 x 17 = 119 minutes, or about 0.5 background events per hour.  Thus, selecting only events having $E > 12 MeV$ is an extremely powerful background suppressor.  

Recall that on a plot of $1/E^2$ versus neutrino arrival time $t$ events corresponding to a specific neutrino mass lie on a straight line passing through the origin whose slope is inversely proportional to $m^2$.  Given that 8 of the 12 actual SN 1987A events observed in Kamioka have $E >12 MeV$ (see Fig. 2), it would not be surprising that any superluminal eigenstate might have perhaps 4 neutrinos associated with it for which $E> 12 MeV.$   The four dots in Fig. 2 (upper) shows what the signature might look like for such a ${m_{\nu_3}}^2 \approx -0.2 keV^2$ signal.  Those four simulated events were arbitrarily assumed to have a uniform time distribution (equivalent to $\frac{dN}{dE}\propto 1/E^3$).  Note that there is no real need to do anything more realistic here, given the extremely low background -- perhaps 0.5 background events in the 1 hour interval before 7:35 UT, as long as we focus only on events having $E>12 MeV.$   Thus,  should the $1/E^2$ versus t plot of the Kamioka neutrino data for this one hour period show perhaps 4 real events falling $\emph{anywhere}$ on a line through the origin having the predicted approximate slope corresponding to ${m_{\nu_3}}^2 \approx -0.2 keV^2,$ this would constitute an unambiguous signature of a superluminal neutrino.

A tantalizing hint that this possibility might be due to more than the author's feverish imagination is provided by the one real Kamioka event (the square in Fig. 2 upper graph) that falls in the only 17 min time interval falling in the one hour before the 12 event burst.  This event lies quite near the predicted straight line.  It is silly to provide a calculation of the probability of this occurring based on random background, given only one event, but it is probably only about 1/100.  This estimate is based on the likelihood of a background event occuring in that 17 minute interval (about 1/10), and its likelihood of it lying very close to the predicted line (about 1/10).  This one event proves nothing, but if it were possible for Kamioka, IMB and Baksan to reexamine their old data one might find more persuasive confirming evidence.    Thus, the probability of N events falling on or near the line as a result of random background would be on the order of $p = 10^{-2N}.$

\newpage

\begin{figure}[h]
\epsfxsize=200pt 
\epsffile{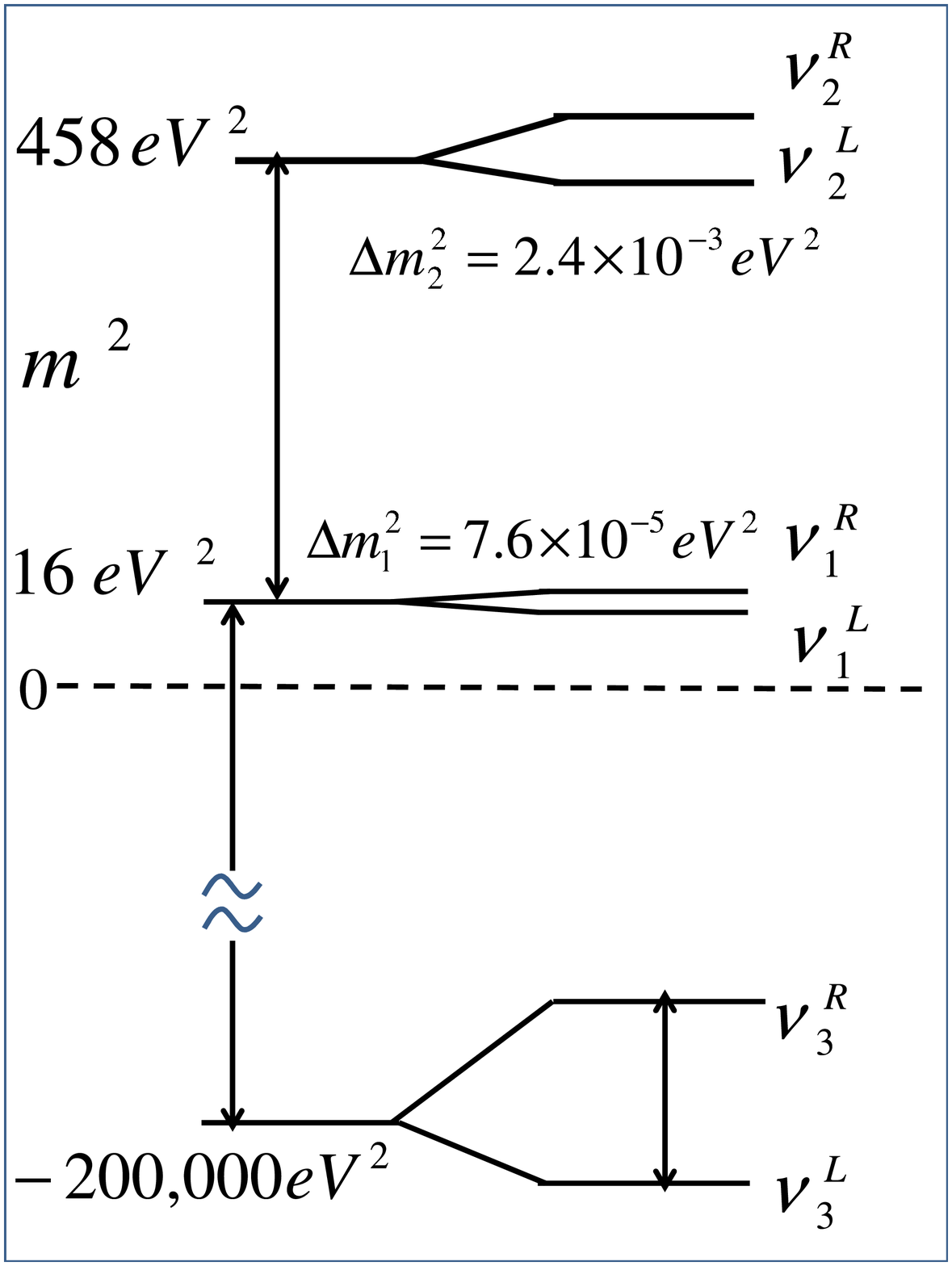} \\
\caption{\small A 3 + 3  neutrino model involving three active and three sterile neutrinos arranged in nearly degenerate active-sterile pairs, with the third pair having ${m_3}^2<0.$  The two mass splittings labelled $\Delta m^2_1$ and $\Delta m^2_2$ are the well-measured values normally labelled $\Delta m^2_{1,2}$ and $\Delta m^2_{2,3}$ found in neutrino oscillation experiments.  The double arrows show the only three independent large mass differences in this model, with the shortest double arrow assumed to correspond to $\Delta m^2_3\approx 1 eV^2.$  The correspondence between our labelling of the six neutrino mass states and the conventional one is: $\nu_1=\nu_{1L},\hspace{0.25in} \nu_2=\nu_{2L}, \hspace{0.25in} \nu_3=\nu_{3L}, \hspace{0.25in}\nu_4=\nu_{1R},\hspace{0.25in} \nu_5=\nu_{2R}, \hspace{0.25in} \nu_6=\nu_{3R}.$ 
 \\}
\end{figure}

\begin{figure}[h]
\epsfxsize=250pt 
\epsffile{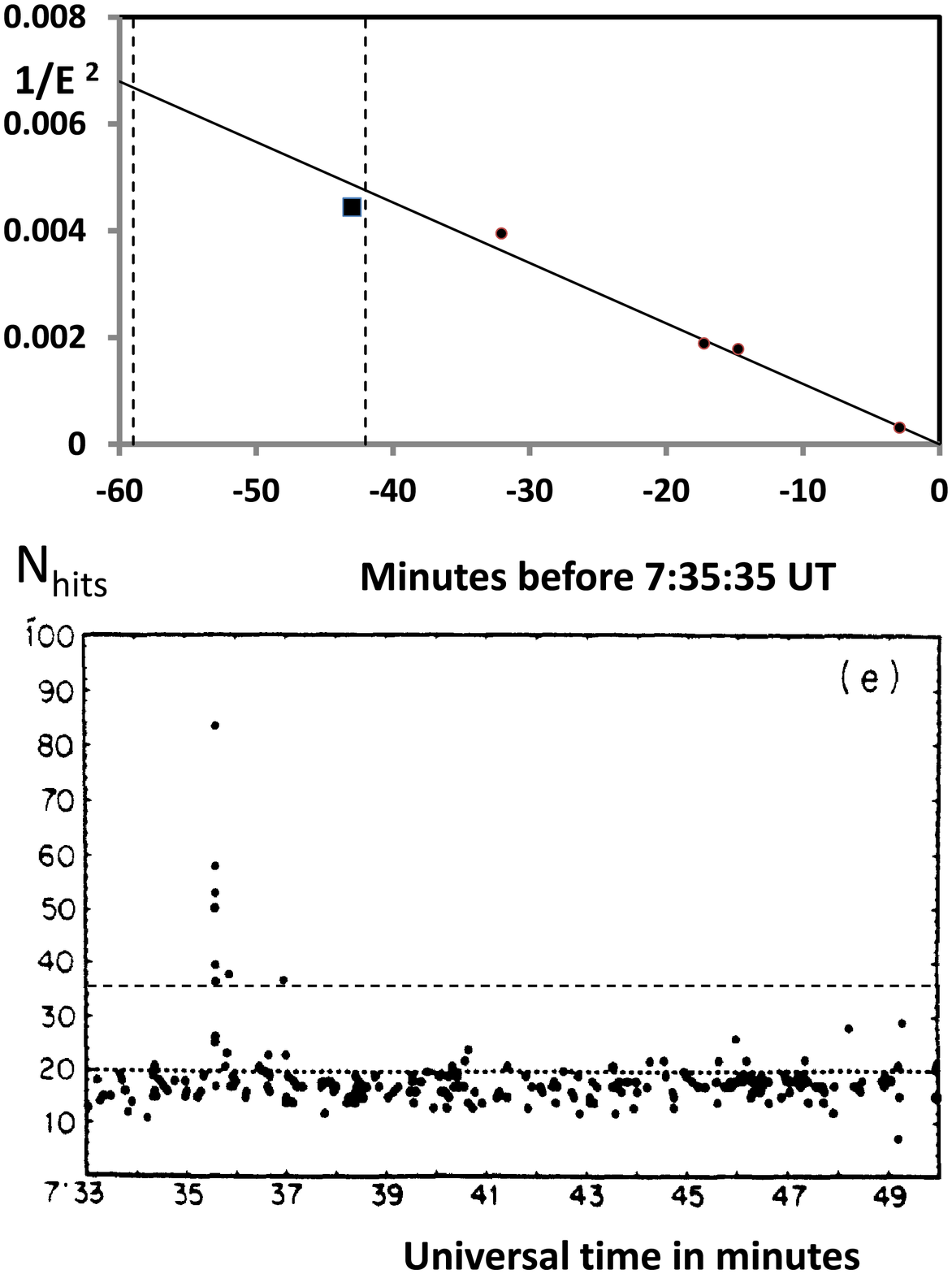} \\
\caption{\small \emph{Lower graph}:  Fig. 4 (e) in Hirata et. al.\cite{Kamioka} showing the 12 event burst from SN 1987A.  $N_{hits}$ is a measure of the neutrino energy, and 35 hits corresponds to $E_{\nu} \approx 12 MeV.$  \emph{Upper graph}:  Plot of $1/E^2$ in $MeV^{-2}$ for four randomly generated \emph{simulated} events shown as small dots corresponding to a ${m_{\nu_3}}^2 \approx -0.2 keV^2$ signal.  The point labeled by a large square is the only real event having $E > 12 MeV$  in the 17 minute interval defined by the two vertical lines.
 \\}
\end{figure}

\begin{thebibliography}{00}

\bibitem{Bilaniuk}  Bilaniuk, O.-M. P.; Deshpande, V. K.; Sudarshan, E. C. G. "Meta' Relativity". American Journal of Physics 30, 718 (1962)

\bibitem{Chodos} A. Chodos, A. Hauser and V. A. Kostelecky, Phys. Lett. B 150, 431 (1985)

\bibitem{Bertolucci} S. Bertolucci, presentation at Neutrino 2012 in Kyoto, on behalf of the Borexino, ICARUS, LVD, and OPERA collaborations, June 8, 2012.

\bibitem{Adamson} P.A. Adamson, presentation at Neutrino 2012 in Kyoto, on behalf of the MINOS collaborations, June 7, 2012.

\bibitem{Kamioka} K. Hirata, et. al., "Observation of a neutrino burst in coincidence with supernova 1987A in the Large Magellanic Cloud,"  Phys. Rev. Lett. 58, 1490-1493 (1987)

\bibitem{IMB} R. M. Bionta et. al., "Observation of a neutrino burst in coincidence with supernova 1987A in the Large Magellanic Cloud,"  Phys. Rev. Lett. 58, 1494-1496 (1987)

\bibitem{Baksan} E. N. Alekseev, L. N.  Alekseeva, I. V. Krivosheina, V. I. and Volchenko, "Detection of the neutrino signal from SN 1987A using the INR Baksan underground scintillation telescope," ESO Workshop on the SN 1987A, Garching, Federal Republic of Germany, July 6-8, 1987, Proceedings (A88-35301 14-90). Garching, Federal Republic of Germany, European Southern Observatory, 237-247 (1987)

\bibitem{Mt Blanc} M. Aglietta, et. al., "On the Event Observed in the Mont Blanc Underground Neutrino Observatory during the Occurrence of Supernova 1987a," Europhys. Lett. 3 (12) 1315.

\bibitem{Pagliaroli} See extensive list of references in: G. Pagliaroli, F. Vissani, M. L. Costantini, and A. Ianni1, "Improved analysis of SN1987A antineutrino events," Astroparticle Phys., 31, (3), 163-176 (2009)

\bibitem{Bahcall} D. N. Spergel, J. N. Bahcall,  "The mass of the electron neutrino: Monte Carlo studies of SN 1987A observations," Phys. Lett. B, 200, (3), 366-372 (1988)

\bibitem{Arnett} W. D. Arnett, and L.J. Rosner.,  "Neutrino mass limits from SN1987A,"  Phys. Rev. Lett.,   58, 1906-1909  (1987).

\bibitem{Pagliaroli1} G. Pagliaroli, F. Rossi-Torres, F. Vissani, "Neutrino mass bound in the standard scenario for supernova electronic antineutrino emission,” Astropart. Phys. 35, 287-291 (2010) 

\bibitem{Ellis} Ellis, J. et. al., "Prospective constraints on neutrino masses from a core-collapse supernova," Phys. Rev. D, 85 (10)1-6 (2012)

\bibitem{Ehrlich0} R. Ehrlich,  “Evidence for two neutrino mass eigenstates from SN 1987A and the possibility of superluminal neutrinos,” Astropart. Phys. 35, (10) 625-628 (2012)

\bibitem{Huzita}  H. Huzita, "Neutrino mass speculation on the neutrino events from the supernova LMC 1987 A," Huzita, H., Mod. Phys. Lett. A2 (1987) 905-911.

\bibitem{Cowsik} R. Cowsik, "Neutrino masses and flavors emitted in the supernova SN1987A," Phys. Rev. D 37, 1685–1687 (1988)

\bibitem{Totani} T.Totani, K.Sato, H.E. Dahled, and J.R. Wilson, Future detection of supernova neutrino burst and explosion mechanism, Ap.J., 496, 216-225 (1998)

\bibitem{Bruenn} S.W. Bruenn, "Neutrinos from SN1987A and current models of stellar-core collapse, Phys. Rev. Lett. 59, 938–941 (1987)

\bibitem{Myra} E.S. Myra, A. Burrows, "Neutrinos from type II supernovae - The first 100 milliseconds," Ap.J. 364, 222-231 (1990)

\bibitem{Lamb} T. J. Loredo, and D. Q. Lamb, "Bayesian analysis of neutrinos observed from supernova SN 1987A," Phys. Rev. D, 65, 063002 (2002)

\bibitem{Janka} Thomas Janka, Private communication.

\bibitem{Raffelt} G. G. Raffelt, "Physics opportunities with supernova neutrinos," Progress in Particle and Nuclear Physics, 64, 393–399 (2010)


\bibitem{Thomas} S. A. Thomas, F. B. Abdalla, and O. Lahav, Phys. Rev. Lett. 105, 031301 (2010)

\bibitem{PDG} J. Beringer et al. (Particle Data Group), J. Phys. D86, 010001 (2012) 

\bibitem{Kopp} J. Kopp, M. Maltoni, T. Schwetz, Phys. Rev. Lett. 107, 091801 (2011)

\bibitem{Conrad} J. M. Conrad, C.M. Ignarra, G. Karagiorgi, M.H. Shaevitz, J. Spitz, "Sterile Neutrino Fits to Short Baseline Neutrino Oscillation Measurements," to appear in a special review article in Neutrino Physics,  arXiv:1207.4765

\bibitem{Berezhiani} Z. G. Berezhiani and R.N. Mohapatra, "Reconciling present neutrino puzzles: Sterile neutrinos as mirror neutrinos," Phys. Rev. D 52, 6607–6611 (1995)

\bibitem{Louis} W. Louis, "Particle physics: Sterile neutrinos," Nature 478, 328–329 (2011)
 
\bibitem{Kostelecky} V. A. Kostelecky, in F. Mansouri, J.J. Scanio (Eds.), Topics in Quantum Gravity and Beyond, World Scientific, Singapore, 1993.

\bibitem{Ehrlich1} R. Ehrlich, "Implications for the Cosmic Ray Spectrum of a Negative Electron Neutrino Mass$^2$," Phys. Rev. D, 60, 17302 (1999)

\bibitem{Ehrlich2} R. Ehrlich, "Neutrino mass$^2$  inferred from the cosmic ray spectrum and tritium beta decay, Physics Letters B 493 (2000) 229-232

\bibitem{Ehrlich3} R. Ehrlich, "Is There a 4.5 PeV Neutron Line in the Cosmic Ray Spectrum?" Phys. Rev. D, 60, 73005 (1999)

\bibitem{r14} J. Lloyd-Evans et. al., "Observation of γ rays $>10^{15} eV$ from Cygnus X-3." Nature 305, 784 - 787 (27 October 1983)

\bibitem{r13} A. Borione et al., "High Statistics Search for Ultrahigh Energy Gamma-Ray Emission from Cygnus X-3 and Hercules X-1," Phys. Rev. D, 55 (1997), 1714.

\bibitem{Ciborowski} J. Ciborowski, J. Rembielinski, "Tritium Decay and the Hypothesis of Tachyonic Neutrinos, Eur. Phys. J. C8:157-161 (1999)

\bibitem{Radzikowski} M. J. Radzikowski,  "A quantum field model for tachyonic neutrinos with Lorentz symmetry breaking," arXiv:1007.5418 (July 2010)

\bibitem{Cowsik1}  R. Cowsik, S. Nussinov, and U. Sarkar, "Superluminal Neutrinos at OPERA Confront Pion Decay Kinematics," Phys. Rev. Lett., 107, 251801 (2011)

\bibitem{Cohen} A. G. Cohen, S. L. Glashow, " New Constraints on Neutrino Velocities," arXiv:1109.6562v1

\bibitem{Miniboone}  The Miniboone Collaboration,  A. A. Aguilar-Arevalo, et. al, "Test of Lorentz and CPT violation with Short Baseline Neutrino Oscillation Excesses" submitted to Phys. Lett. B, arXiv:1109.3480

\bibitem{Adamson1} P. Adamson, et. al., "Search for Lorentz invariance and CPT violation with muon antineutrinos in the MINOS Near Detector,"  arXiv:1201.2631

\bibitem{Icecube} IceCube Collaboration: R. Abbasi, et. al., "Search for a Lorentz-violating sidereal signal with atmospheric neutrinos in IceCube," Phys Rev D.82.112003

\bibitem{Kostelecky2} D. Colladay and V. A. Kostelecky, "Lorentz-violating extension of the standard model," Phys. Rev. D 58, 116002 (1998)

\bibitem{Chodos2} A. Chodos, "Light Cone Reflection and the Spectrum of Neutrinos," arXiv:1206.5974 (June 2012)

\bibitem{Jentschura}  U. D. Jentschura, B. J. Wundt, "Localizability of tachyonic particles and neutrinoless double beta decay,  The Eur. Phys. Journal C, 72, 1894 (2012) 

\bibitem{Jentschura2} Private informal communication.

\bibitem{Jentschura1}  U. D. Jentschura, B. J. Wundt, "From Generalized Dirac Equations to a Candidate for Dark Energy," arXiv:1205.0521

\bibitem{Katrin} K. Valerius for the KATRIN collaboration, "Systematics and background suppression in the KATRIN
experiment, arXiv:0710.4906

\bibitem{Kostelecky3} A. Kostelecky, M. Mewes, "Neutrinos with Lorentz-violating operators of arbitrary dimension,"  Phys. Rev. D 85, 096005 (2012)	arXiv:1112.6395

\end{thebibliography}
\end{document}